\documentclass[preprint2]{emulateapj-rtx4}

\usepackage{txfonts} 
\usepackage{epsfig} 
\usepackage{graphicx}
\usepackage{natbib} 
\usepackage{xspace}
\usepackage{color}
\usepackage[colorlinks,urlcolor=blue,citecolor=blue,linkcolor=blue]{hyperref} 

\tabletypesize{\scriptsize}

\newcommand{\dechms}[4]{$#1^{\rm h}#2^{\rm m}#3\mbox{$^{\rm s}\mskip-7.6mu.\,$}#4$}

\newcommand{\decdms}[4]{$+#1^{\circ}#2'#3\mbox{$''\mskip-7.6mu.\,$}#4$}

\slugcomment{ApJ}

\shorttitle{Molecular outflows: Explosive versus protostellar}
\shortauthors{Zapata et al.}

\begin{document}


\title{Molecular outflows: Explosive versus protostellar}


\author{Luis A. Zapata\altaffilmark{1}, Johannes Schmid-Burgk\altaffilmark{2},\\  
Luis F.  Rodr\'\i guez\altaffilmark{1}, Aina Palau\altaffilmark{1}, and Laurent Loinard\altaffilmark{1}}


\altaffiltext{1}{Instituto de Radioastronom\'\i a y Astrof\'\i sica, UNAM, Apdo. Postal 3-72
  (Xangari), 58089 Morelia, Michoac\'an, M\'exico}
\altaffiltext{2}{Max-Planck-Institut f\"{u}r Radioastronomie, Auf dem H\"ugel	
  69, 53121, Bonn, Germany}


\begin{abstract}
With the recent recognition of a second, distinctive class of molecular outflows, namely the explosive ones 
not directly connected to the accretion-ejection process in the star formation, a juxtaposition of the morphological 
and kinematic properties of both classes is warranted. By applying the same method used in \citet{zap2009}, and using $^{12}$CO(J=2-1) 
archival data from the Submillimeter Array (SMA),
we contrast two well known explosive objects, Orion KL and DR21, to HH211 and DG Tau B, two flows representative 
of classical low-mass protostellar outflows. At the moment 
there are only two well established cases of explosive outflows, but with the full availability of ALMA we expect that more 
examples will be found in the near future.
Main results are the largely different spatial distributions of the explosive flows, consisting of numerous
narrow straight filament-like ejections with different orientations and in almost an isotropic configuration,
the red with respect to the blueshifted components of the flows (maximally separated in protostellar, largely overlapping in explosive outflows),
the very well-defined Hubble flow-like increase of velocity with distance from the origin in the explosive filaments versus the mostly non-organized CO velocity field in
protostellar objects, and huge inequalities in mass, momentum and energy of the two classes, at least for the 
case of low-mass flows. Finally, all the molecular filaments in the explosive outflows point back to approximately a central position 
({\it i.e.} the place where its ``exciting source" was located), 
contrary to the bulk of the molecular material within the protostellar outflows.  
\end{abstract}





\section{Introduction}
Since their discovery, 
the explosive outflows have formed part of a new class of molecular outflows related with star forming regions \citep{zap2009}.  
At this moment, there seem to exist two kinds of molecular outflows related with star forming regions \citep{fra2014,ball2016}. 
One kind are the {\it protostellar} molecular outflows energized by a star in its process of formation 
\citep[see for a review,][]{arce2007}, and the other kind are the {\it explosive} molecular outflows, probably associated
with the disruption of a non-hierarchical massive and young stellar system (triggered by the possible merger of young massive stars), 
or with a protostellar collision \citep[{\it e.g.},][]{zap2009,ball2011,zap2013, riv2014,ball2016,ball2017}. 
Hence,  the occurrence of the two types of molecular flows could point to the existence of two totally different physical phenomena \citep[][]{zap2009}. 
In order to distinguish between the two outflow phenomena it is therefore required to study their morphological and 
kinematic differences. 

The general agreement is that the molecular outflows associated with the process of star formation 
(energized by means of a circumstellar accreting disk) are probably formed by the interaction between the molecular 
material from the parental cloud and the ionized/neutral collimated jet ejected from the young star, that creates internal jet working surfaces \citep{mas1993,rag1993}.  
This interaction additionally produces bow-shocks, cavities, swept-up shells and jet shocks, see for example 
the cases of the HH 211, DG Tau B,  IRAS 18162-2048-NW and IRAS 18360-0537
\citep{fro2005, palau2006, qiu2012, man2013, zap2015} or the Serpens South outflow \citep{plu2015}. 
However, there is a large amount of cases where even molecular material is present in form of collimated jets 
(where the ionized/neutral gas is supposedly located), 
see for an example the cases of the HH 211 or Serpens South outflows \citep{palau2006, plu2015}. 
The nature of the primary collimated jet is still under debate.  At this time, it is hypothesized that such primary jets
 are driven and collimated by rapidly rotating magnetic fields coupled to the star/disk system \citep{pud2007, fra2014, arce2007}.    
  
On the other hand are the explosive outflows, recently reported and discussed in the literature \citep[{\it e.g.},][]{zap2009,zap2013,ball2011,you2016,ball2016}.  
The explosive flows seem to be impulsive, and created possibly by an energetic single and brief event \citep{ball2005}.
These flows consist of dozens of expansive CO filaments, [FeII] fingers, and H$_2$ wakes pointing back approximately to a central position, reminiscent 
of an explosive event, see the cases of DR21 or Orion KL \citep[][]{alle1993,zap2009,zap2013, you2016}. 
The  expansive CO filaments are nearly isotropic on the sky and present well defined Hubble velocity laws, that is the radial velocities increase 
with the projected distances \citep[see Figure 2 of][]{zap2009,zap2013,ball2017}.  In the case of Orion KL, the dynamical ages of most CO filaments
are close to 500 yrs, the age of the disruption of a non-hierarchical massive and young stellar system \citep{Gom05,zap2009} which is the event that probably originated the outflow. 
As the explosive flow in Orion KL has no longer an ``exciting" source it should not last for a long time as compared to the long-lifes of the 
protostellar outflows with dynamical timescales of about 10$^{4-5}$ years \citep{fra2014,ball2016}.  

The precise nature of the explosive outflows is still not well understood but it is speculated that
it might be related to the disintegration of young stellar systems --as mentioned above and revealed in the case of Orion KL  
\citep[][]{ball2005,zap2009,ball2011,ball2017}. \citet{ball2015} proposed that the optical bullets in Orion KL are probably 
dense fragments of interacting circumstellar disks that were torn apart in
a stellar encounter and launched into the surrounding cloud.  The high tangential velocities of the optical bullets \citep[$\sim$300 km s$^{-1}$: ][]{Doi2002} were inherited from the 
very inner Keplerian velocities of the disks. This implies that the H$_2$ wakes are probably warm gas trails of a large numbers of expelled chunks 
or bullets observed at a wide range of wavelengths. As to the CO filaments, their nature is unclear, they could also be small chunks of disks with different velocities, or
warm gas trails.
      
In this study, we present Submillimeter Array $^{12}$CO(J=2-1) observations of the two classical molecular outflows HH211 and DG Tau B excited by 
stars in their process of formation, and of two recently recognized explosive outflows (Orion KL and DR21) associated with massive star forming regions.
The main reason for this comparison with the low-mass stars is the large outflow multiplicity observed in the high-mass protostars, see for example the cases of IRAS 05358+3543 \citep{beu2002}, G240.31+0.07 \citep{qiu2009}, and IRAS 17233-3606 \citep{leu2009,kla2015}. This high outflow multiplicity difficulties the identification 
of any physical and kinematical difference between both kinds of flows.      
 The new analysis conducted in these CO observations allow us to compare for the first time, the morphological and kinematic properties of both classes of molecular outflows. 
      
\begin{figure*}[!]
\begin{center}
\hspace{-2.2cm}
\includegraphics[scale=0.3]{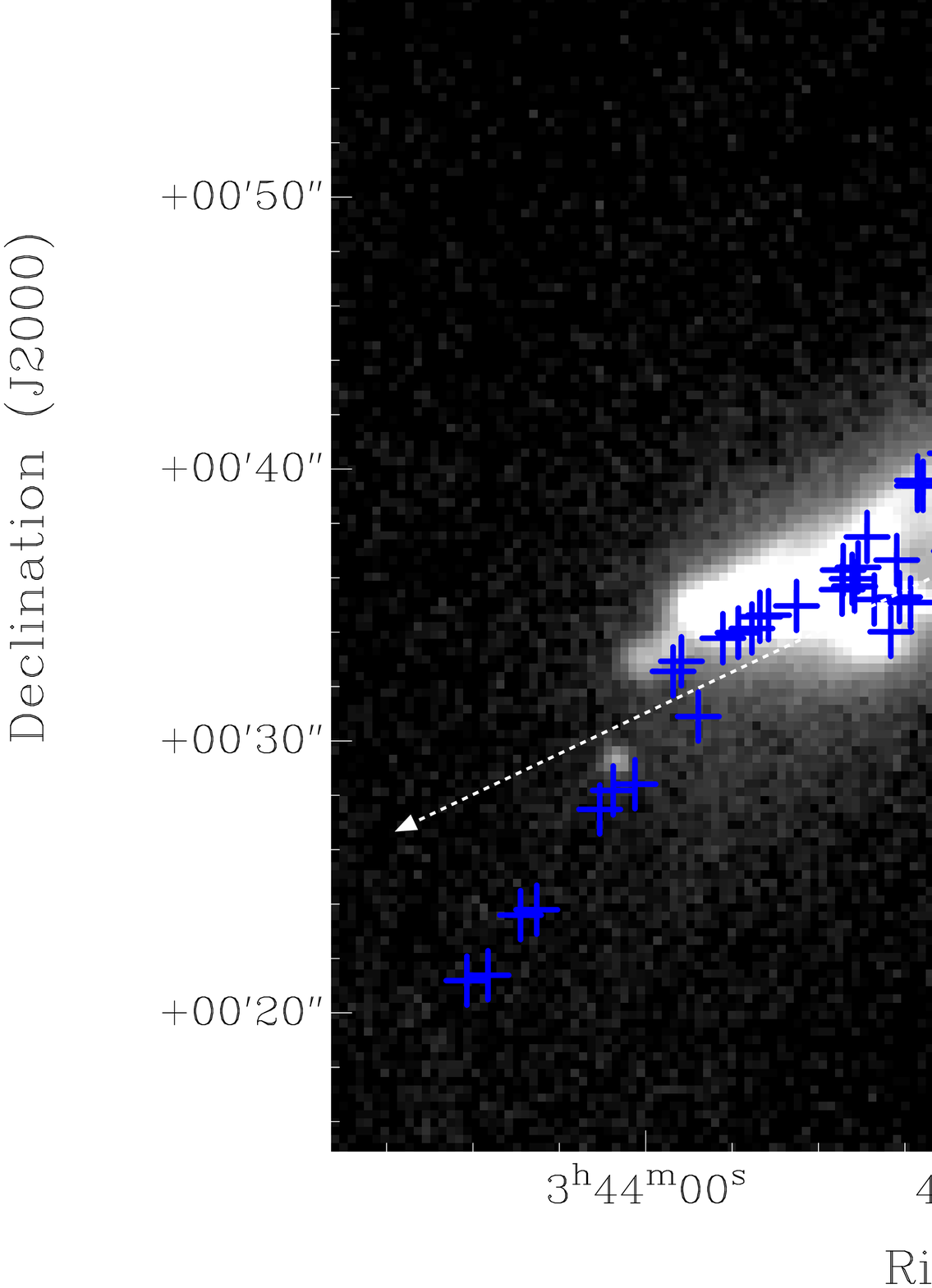}
\includegraphics[scale=0.3]{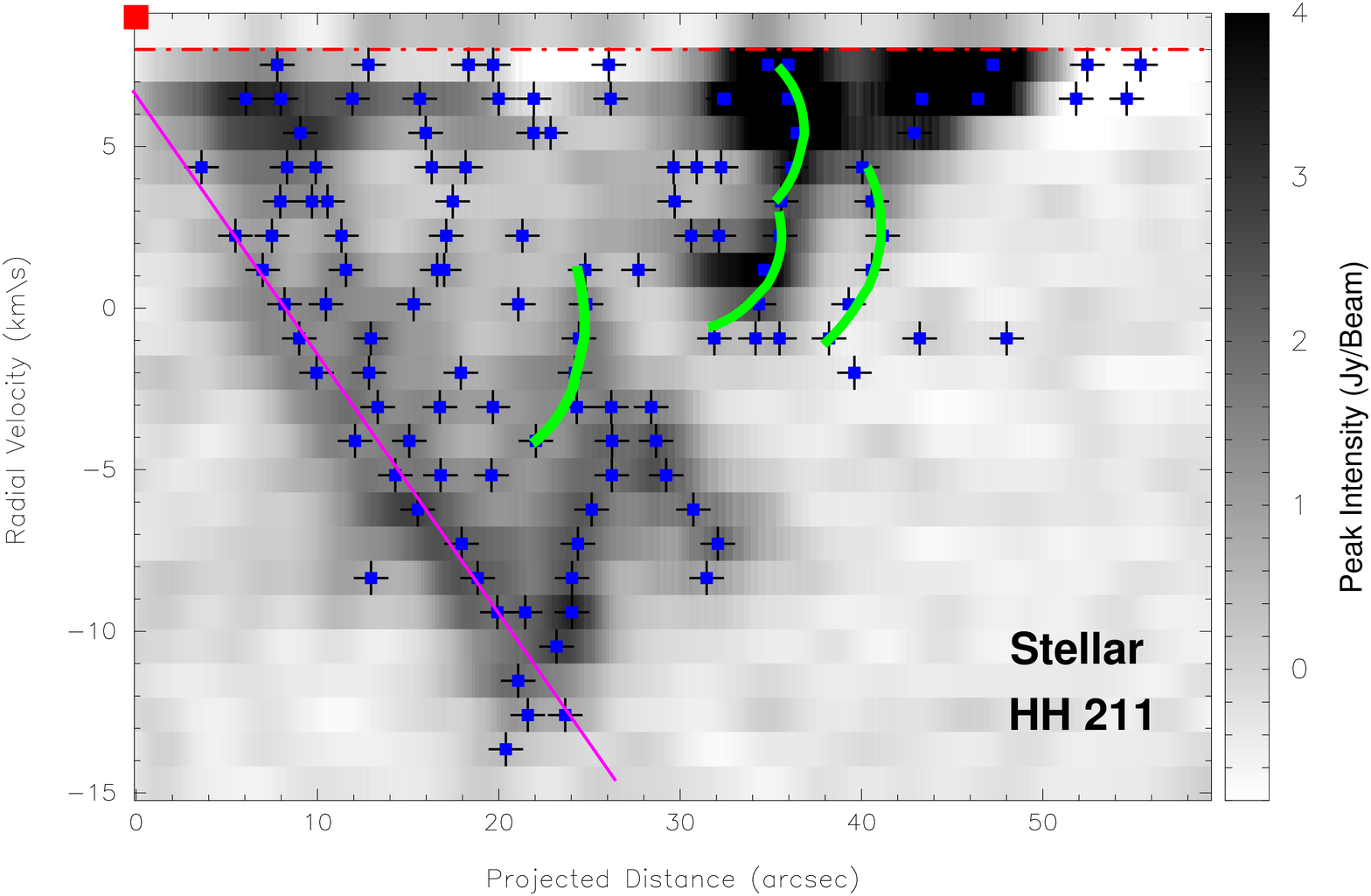}
\caption{\scriptsize  {\it Upper:} H$_2$ infrared line (2.122 $\mu m$) image of the blueshifted side of the molecular outflow HH 211 \citep{mac1994}, 
overlaid with the SMA positions of the $^{12}$CO(J=2-1) 
blueshifted compact condensations revealed in the spectral data cube (blue crosses), and 
the 3.3 cm continuum emission mapped by the VLA \citep[white contours,][]{rod2014}. 
The red circle and blue cross (on top of the white contours) mark the peak position of the 3.3 cm source. 
The white contours are in percent of the peak emission starting from 40\% to 90\%, in steps of 10\%. 
The intensity peak emission is 0.18 mJy Beam$^{-1}$. The half-power contour of the synthesized beam 
of the 3.3 cm image is shown in the bottom right corner.  
The dashed arrow marks the position of the cut shown below and its orientation 
toward positive values.
{\it Lower:} Position velocity diagram of the blueshifted $^{12}$CO(J=2-1) emission from the HH 211 outflow, overlaid 
with the SMA positions of the blueshifted compact condensations revealed in each channel of the spectral data cube (blue crosses).  The position velocity diagram
is made along the blueshifted jet axis at a P.A. = 115$^\circ$, see upper panel.  
peaks revealed in the position-velocity diagram. The red dashed line marks the systemic velocity of the cloud. The emission inside of this range cannot be well sampled by the SMA. 
The red square indicates the position of the continuum 3.3 cm source and its systemic velocity \citep[$\sim$9.0 km s$^{-1}$;][]{palau2006}.
The pink line marks the tentative Hubble velocity law observed in the protostellar outflow. The green curve trace the bow-shocks revealed across the flow. }
\label{fig1}
\end{center}
\end{figure*}

\begin{figure*}[t]
\begin{center}
\hspace{-2.0cm}
\includegraphics[scale=0.3]{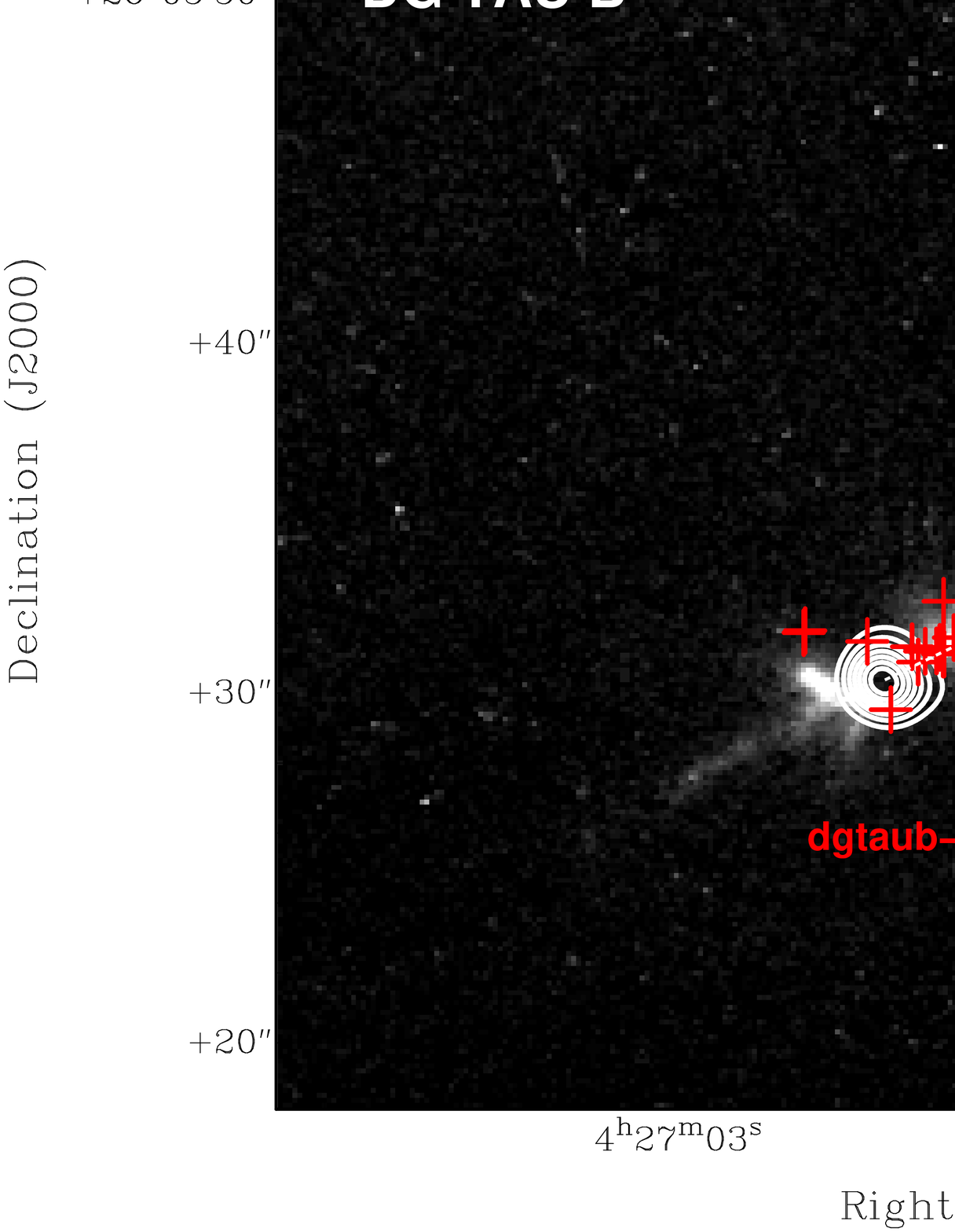}
\includegraphics[scale=0.3]{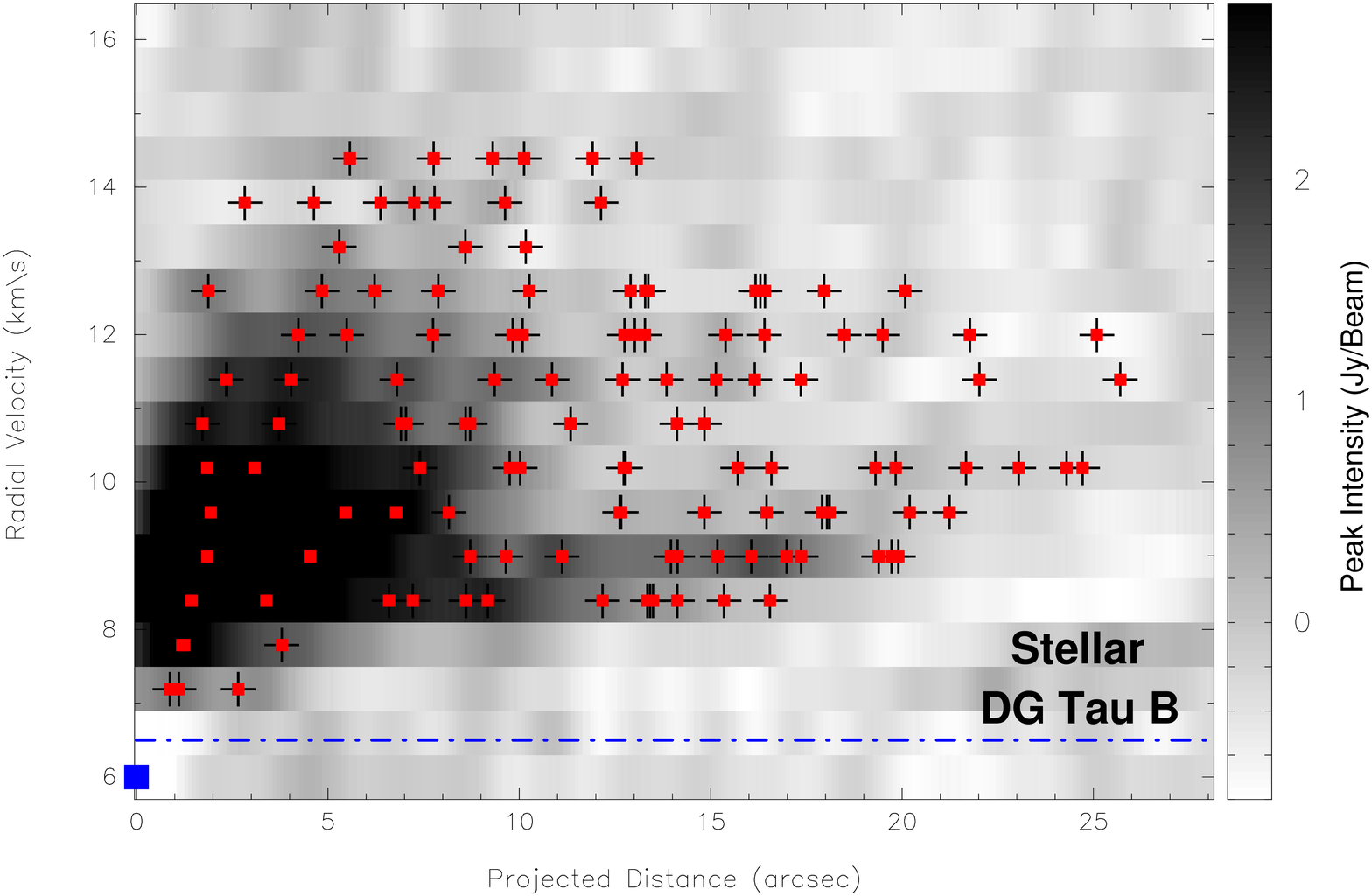}
\caption{\scriptsize {\it Upper: } HST-WFPC optical image (F675w) of the molecular outflow DG Tau B, 
overlaid with the SMA positions of the $^{12}$CO(J=2-1) 
redshifted compact condensations revealed in the spectral data cube (red crosses), and 
the 1.3 mm continuum emission mapped with the SMA \citep[white contours,][]{zap2015}. 
The white contours are in percent of the peak emission starting from 30\% to 90\%, in steps of 10\%. 
The intensity peak emission is 0.31 Jy Beam$^{-1}$. The half-power contour of the synthesized beam 
of the 1.3 mm image is shown in the bottom right corner.  
The dashed arrow marks the position of the cut shown below and its orientation 
toward the positive values.
{\it Lower: } Position-velocity diagram of the redshifted $^{12}$CO(J=2-1) emission from the DG Tau B outflow, overlaid 
with the SMA positions of the redshifted emission peaks in each channel revealed in the spectral data cube (red crosses).  The PV diagram
is made along the red-shifted jet axis at a P.A. = 296$^\circ$, see Figure 3.  
the systemic velocity of the cloud. The emission inside of this range cannot be sampled well by the SMA. 
The blue square indicates the position of the continuum 1.3 cm source and its systemic velocity \citep[$\sim$6.5 km s$^{-1}$;][]{zap2015}.}
\label{fig2}
\end{center}
\end{figure*}

\section{Archival SMA Observations}

\subsection{HH211}

The millimeter $^{12}$CO(J=2-1) data were collected on 2008 November 16 with the Submillimeter Array\footnote{The Submillimeter Array (SMA) 
is a joint project between the Smithsonian Astrophysical Observatory and the Academia Sinica Institute of Astronomy and Astrophysics, and is funded by the Smithsonian 
Institution and the Academia Sinica.},  when the array was in its compact configuration. The independent baselines in this
configuration ranged in projected length from 6 to 60 k$\lambda$. The phase reference center of the observations was situated
at the position $\alpha_{J2000.0}$ = \dechms{04}{27}{02}{66}, $\delta_{J2000.0}$ = \decdms{26}{05}{30}{4}, located toward the southeastern 
terminal shock of the HH211 jet. This allowed to better trace the molecular component from this side of the flow. 
The SMA primary beam is approximately 55$''$ at 230 GHz. During these millimeter observations the zenith opacity was reasonable, 
ranging from 0.1 to 0.3, and resulting thus in stable phases during the whole track. The digital correlator was set to have 24 spectral "chunks" of 104 MHz 
and 256 channels each. This spectral resolution yielded a velocity resolution of about 0.5 km s$^{-1}$ per channel, however, we smoothed it to $\sim$ 1.0 km s$^{-1}$. 
The quasars 3C84, and 3C111 were used as bandpass and gain calibrators, while one of the Saturn's moon, Titan, was used as flux calibrator. 
The uncertainty in the flux scale is estimated to be between 15\% and 20\%, based on the SMA monitoring of quasars. The rest frequency of the $^{12}$CO(J=2-1)
is  230.5379944 GHz and was detected in the upper sideband of the SMA observations.
 
The $^{12}$CO data were calibrated using the IDL superset {\emph MIR} adapted for the 
SMA\footnote{The MIR-IDL cookbook by C. Qi can be found at http://cfa-www.harvard.edu/~cqi/mircook.html}. 
The calibrated data were imaged and analyzed in the standard manner using the {\emph MIRIAD} \citep{sau1995} and
{\emph KARMA} \citep{goo96} softwares\footnote{The raw data can be obtained from: 
http://www.cfa.harvard.edu/}. A $^{12}$CO(J=2-1) velocity cube was obtained setting the {ROBUST} 
parameter of the task {INVERT} to $+$2 to obtain a better sensitivity while losing some angular resolution. The contribution from the continuum was removed.
The resulting r.m.s.\ noise  for the velocity data cube was about  30 mJy beam$^{-1}$ per channel, 
at an angular resolution of $3\rlap.{''}0$ $\times$ $2\rlap.{''}4$ with a P.A. = $-56.7^\circ$.

\subsection{DG Tau B}

The millimeter $^{12}$CO(J=2-1) data were collected on 2011 November with the Submillimeter Array,  when the array was in its compact configuration. 
The independent baselines in these configurations ranged in projected length from 10 to 140 k$\lambda$.
The phase center is located at the position $\alpha_{J2000.0}$ = \dechms{04}{27}{02}{66}, $\delta_{J2000.0}$ = 
\decdms{26}{05}{30}{4}.  The SMA digital correlator was set to have 48 spectral chunks" of 104 MHz 
and 128 channels each. This spectral resolution yielded a velocity resolution of about 1 km s$^{-1}$ per channel,
however, the spectral "chunk" where the CO is located allowed 512 channels of resolution providing
a better velocity resolution ( $\sim$0.5 km s$^{-1}$).  The planet Uranus was used to obtain the absolute scale for the flux density calibration.
The gain calibrators were the quasars 3C 111 and 3C 84, while 3C 279 was used for bandpass calibration. 

The CO data were calibrated and imaged using {\emph MIR},  {\emph KARMA,} and  {\emph MIRIAD}. 
A CO velocity cube was obtained setting the {ROBUST} 
parameter of the task {INVERT} to $+$2 to obtain a better sensitivity losing some angular resolution. 
The contribution from the continuum was removed.
The resulting r.m.s.\ noise for the cube line was about  90 mJy beam$^{-1}$ per channel, 
at an angular resolution of $3\rlap.{''}28$ $\times$ $2\rlap.{''}9$ with a P.A. = $-64.5^\circ$.

\subsection{Orion KL}

The $^{12}$CO(J=2-1) data were collected on 2007 January and 2009 January with the Submillimeter Array,  
when the array was in its compact and subcompact configurations. The independent baselines in these 
configurations ranged in projected length from 6 to 58 k$\lambda$.   A mosaic with half-power point spacing between field centers covered the entire 
BN/KL outflow, see \citet{zap2009}.
During these millimeter observations the zenith opacity was also reasonable, 
ranging from 0.1 to 0.3.
The SMA digital correlator was set to have 24 spectral windows (``chunks") of 104 MHz each, 
with 256 channels distributed over each spectral window, thus providing a spectral resolution 
of 0.40 MHz (1.05 km s$^{-1}$) per channel. We smoothed the spectral resolution to 5 km s$^{-1}$.
The planet Uranus, and Titan were used to obtain the absolute scale for the flux density calibration.
The gain calibrators were the quasars J0530$+$135,  J0607$-$085 and J0541$-$056. 
The CO line was detected in the in the upper sideband of the SMA observations.

The CO data were calibrated and imaged using {\emph MIR},  {\emph KARMA,} and  {\emph MIRIAD}. 
A CO velocity cube also was obtained setting the {ROBUST} 
parameter of the task {INVERT} to 0 to obtain an optimal compromise between sensitivity and angular resolution. 
The contribution from the continuum was also removed.
The resulting r.m.s.\ noise for the cube line was about  200 mJy beam$^{-1}$ per channel, 
at an angular resolution of $3\rlap.{''}28$ $\times$ $3\rlap.{''}12$ with a P.A. = $-14.0^\circ$.

\subsection{DR21}

The $^{12}$CO(J=2-1) data were collected on 2011 August and 2012 July/August with the Submillimeter Array,  
when the array was in its extended, compact, and subcompact configuration. The independent baselines in these 
configurations ranged in projected length from 7 to 160 k$\lambda$. This observation used the 
mosaicking mode with half-power point spacing between field centers to cover the entire DR21 flow, see \citet{zap2013}.
The SMA correlator was set to have 24 spectral ``chunks" of 104 MHz 
and 128 channels each. This spectral resolution yielded a velocity resolution of about 1.05 km s$^{-1}$.
During these millimeter observations the zenith opacity was reasonable, ranging from 0.1 to 0.3, and resulting thus in stable 
phases during the whole track. The gain calibrators were the emission-line star MWC 349A, and the quasar J2007+404. 
The planet Uranus was used to obtain the absolute scale for the flux density calibration.
The CO line also was detected in the upper sideband of the SMA observations.

The CO data were calibrated and imaged using {\emph MIR},  {\emph KARMA,}  and  {\emph MIRIAD}. 
A CO velocity cube was obtained setting the {\emph ROBUST} 
parameter of the task {\emph INVERT} to 0 to obtain an optimal compromise between sensitivity and angular resolution. 
The contribution from the millimeter continuum was removed.
The resulting r.m.s.\ noise for the cube line was about  100 mJy beam$^{-1}$, 
at an angular resolution of $2\rlap.{''}17$ $\times$ $1\rlap.{''}89$ with a P.A. = $+74.0^\circ$.

\begin{figure*}[!]
\begin{center} \vspace{-1cm}
\includegraphics[angle=90,scale=0.63]{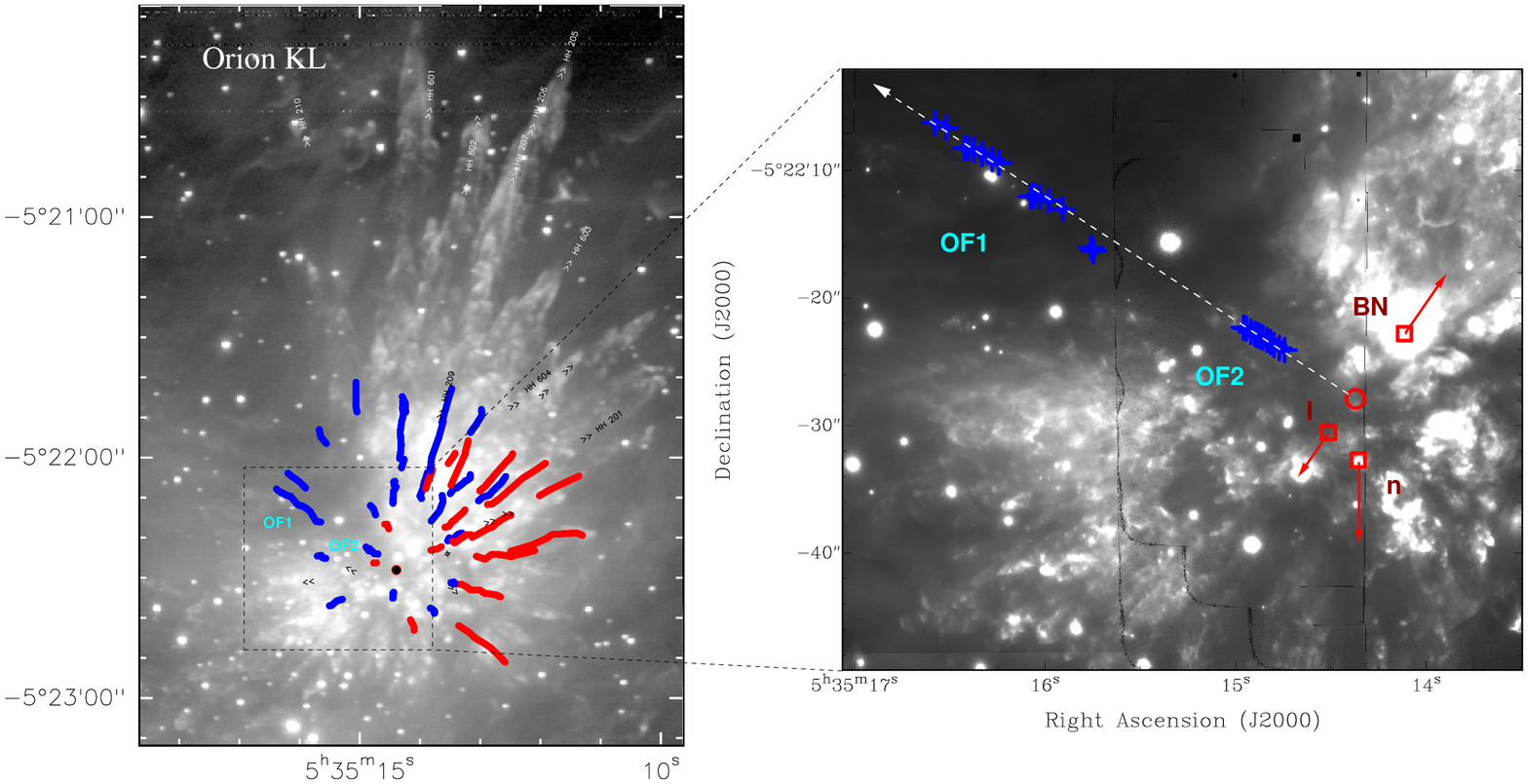}  
\includegraphics[scale=0.35]{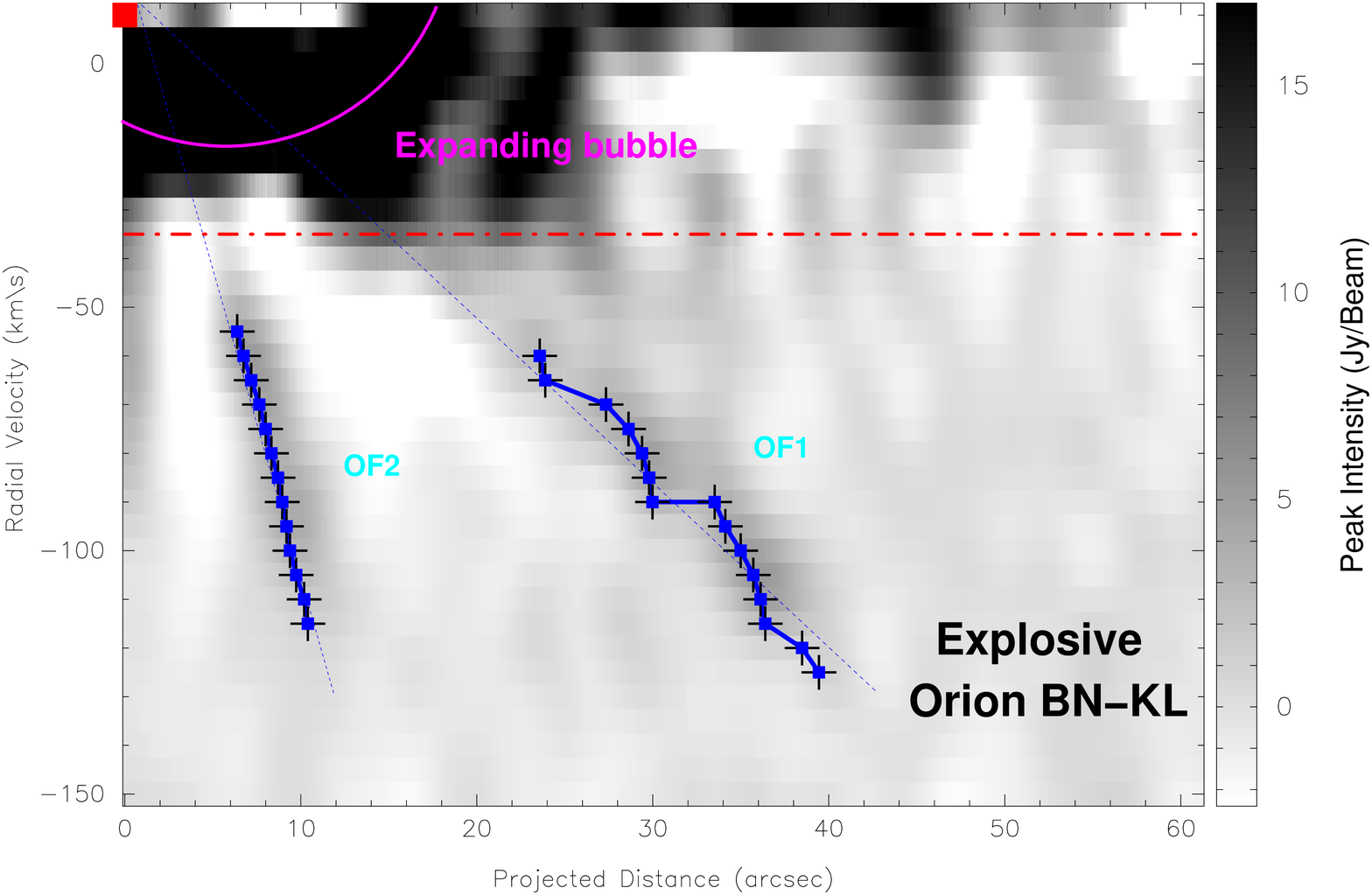}
\caption{\scriptsize  {\it Upper } Left Panel: H$_2$ infrared line image from the molecular outflow in Orion KL  \citep{ball2011,ball2015}, 
overlaid with the SMA positions of the $^{12}$CO(J=2-1) red- and blueshifted compact condensations 
revealed in the spectral data cube (red and blue "filaments") and reported in \citet{zap2009}.
All filaments point toward the same central position marked here with a red/black circle. 
Optical objects moving away from this center are shown as small arrows that indicate the direction 
of their motion \citep{Doi2002,odell2008b}. Right Panel:  a zoom into the center of the outflow overlaid 
with the positions of the runaway sources BN, Source I, and n , and the blueshifted "filaments" OF1 and OF2.  
The red arrows mark the the direction of the proper motion of the three runaway objects
\citep{Rod05,Gom05,Gom08}.
The red circle represents the zone from where the three objects were ejected 
500 years ago \citep{Gom05}.
The dashed arrow marks the position of the cut shown in the lower-panel and its orientation 
toward the positive values. 
 {\it Lower }  Position-velocity diagram of the two blueshifted $^{12}$CO(J=2-1) "filaments" called OF1 and OF2 (shown in the upper panel),   
 overlaid with the positions of the compact condensations revealed in the spectral data cube (blue crosses), see \citet{zap2009}.  
The PV diagram is made along  a P.A. = 54$^\circ$, see  the upper panel.  
The red dashed line marks the position of the starting velocity range where  
the systemic velocity of the cloud is located. The emission inside of this range cannot be well sampled by the SMA. 
The red square indicates the position of the origin of the explosive outflow and its systemic velocity 
\citep[$\sim$9.0 km s$^{-1}$;][]{zap2009,rod2009}. The truncated magenta circle shows part of the expanding molecular 
bubble reported in \citet{zap2011}. The blue dashed lines trace the orientation of each filament. }
\label{fig3}
\end{center}
\end{figure*} 

\begin{figure*}[!]
\begin{center}
\vspace{2cm}
\includegraphics[scale=0.55]{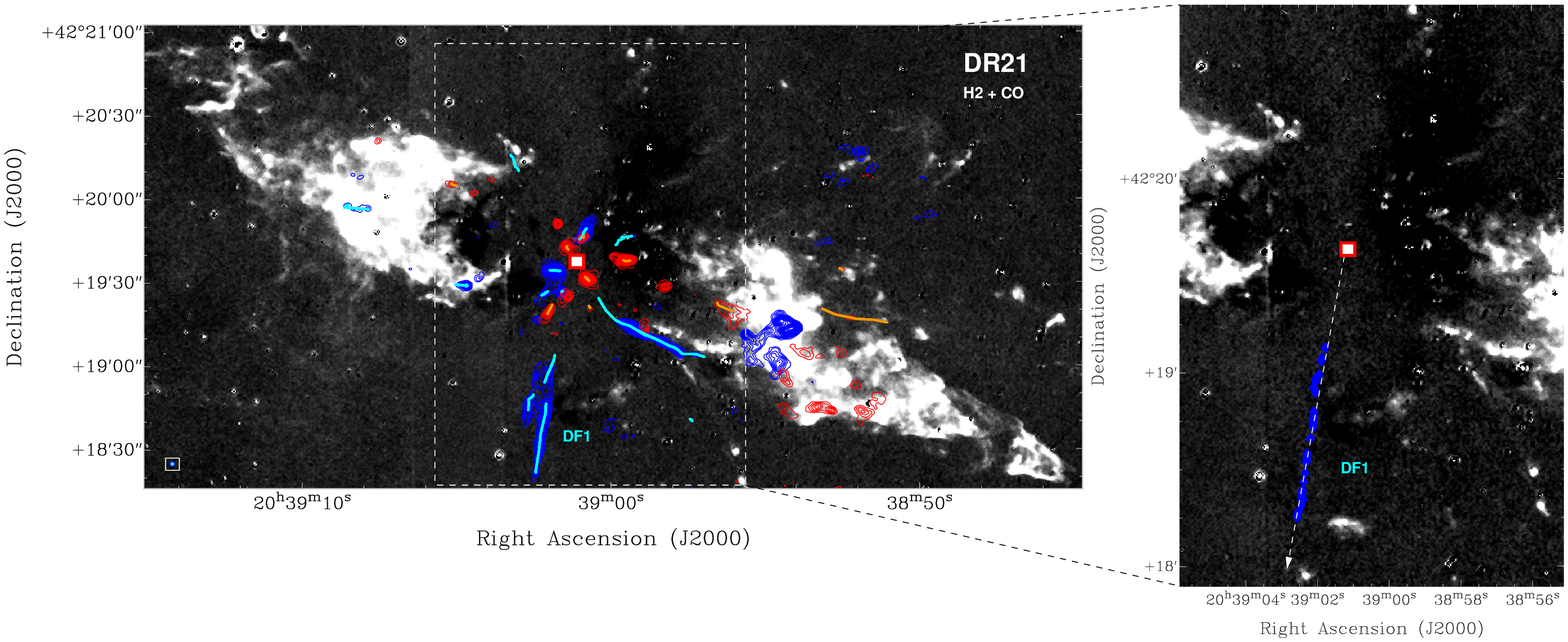}
\includegraphics[scale=0.3]{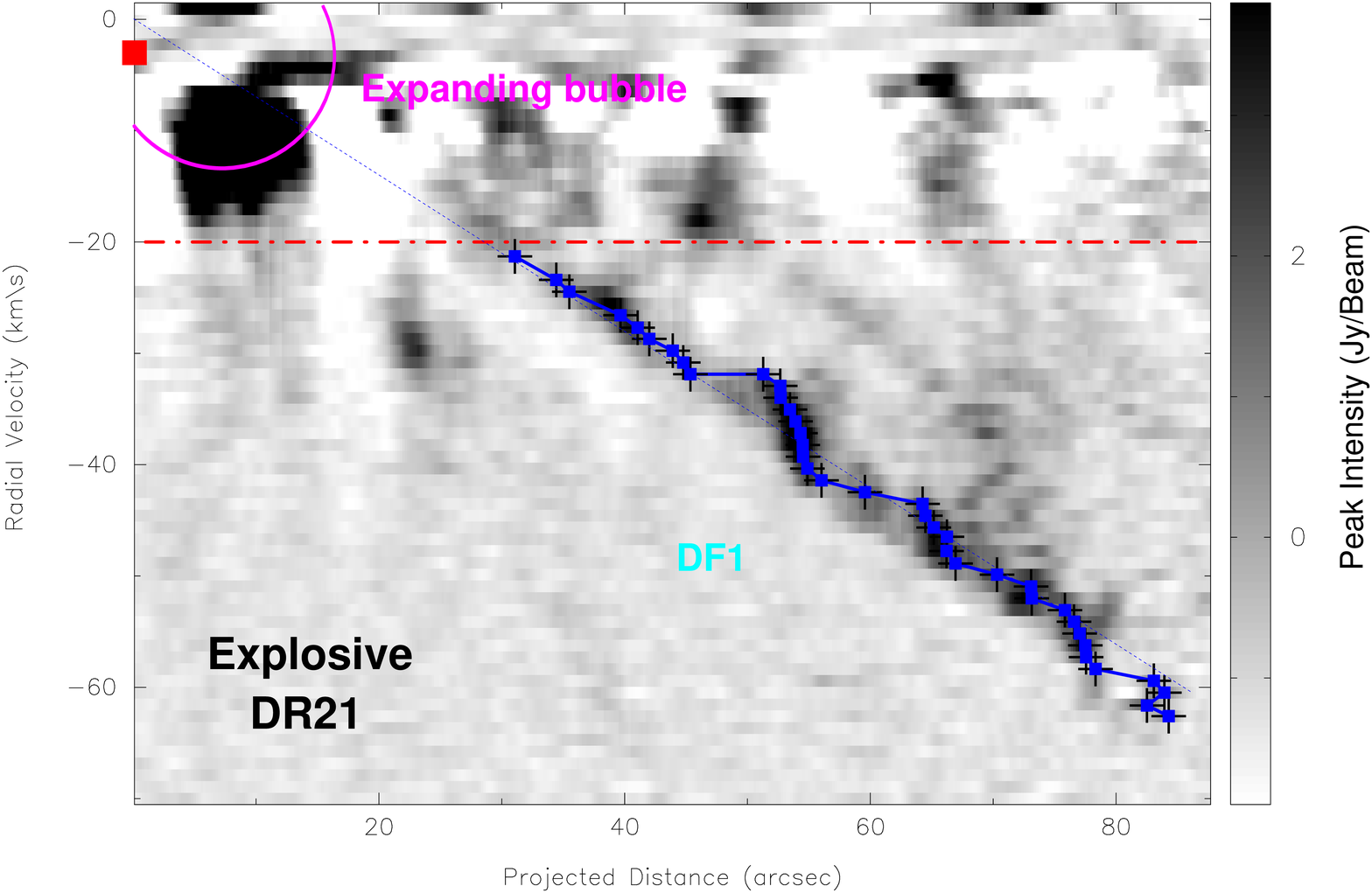}
\caption{\scriptsize {\it Upper:} Left Panel: H$_2$ infrared line image from the molecular outflow in DR21 \citep{dav2007}, 
overlaid with the SMA positions of the $^{12}$CO(J=2-1) red- and blueshifted compact condensations 
revealed in the spectral data cube (red and blue "filaments"), and the moment zero map of the CO emission, reported in \citet{zap2013}.
The blue and red contours trace the $^{12}$CO(J=2-1) molecular emission reported 
in  \citet{zap2013}. The contour values are the same as Figure 1 presented in \citet{zap2013}.
All filaments point toward the same central position marked here with a red box. 
 Right Panel:  a zoom into the center of the outflow overlaid 
with the position of the center of the explosive outflow (red box), and the blueshifted "filament" DF1.  
The dashed arrow marks the position of the cut shown in the lower panel and its orientation 
toward the positive values. 
{\it Lower:} Position-velocity diagram of the blueshifted $^{12}$CO(J=2-1) "filament" called DF1 (shown in the upper-panel),   
 overlaid with the positions of the compact condensations revealed in the spectral data cube (blue crosses), see \citet{zap2013}.  
 The PV diagram is made along  a P.A. = 174$^\circ$, see upper panel.  
 The position of the condensations (shown here as black crosses with a blue square at the center) are in very good agreement with the CO emission 
 peaks revealed in the position-velocity diagram. The red dashed line marks the position of the starting velocity range where  
 the systemic velocity of the cloud is located. The emission inside of this range cannot be well sampled by the SMA. 
 The red square indicates the position of the origin of the explosive outflow and the systemic velocity 
\citep[$-$3.0 km s$^{-1}$;][]{zap2013}. The truncated magenta circle shows part of the expanding molecular 
bubble reported in Zapata et al. (2017), in prep. The blue dashed lines trace the orientation of the filament. }
\label{fig4}
\end{center}
\end{figure*} 

\section{Results}

In this study we have applied the method used in \citet{zap2009} to the spectral data cubes obtained from the SMA archive 
for the outflows: HH211, DG Tau B, Orion KL, and DR 21. In the cases of Orion KL and DR 21, we only applied this to 
a few isolated filaments. This method fits 2D gaussians to extract the position and radial velocities of the molecular condensations
detected above of 5$\sigma$-level above the rms noise of each channel.  

In Figures \ref{fig1}, \ref{fig2}, \ref{fig3}, and \ref{fig4} we show the main results of this study. In these Figures, we present the position of the
most prominent CO features revealed in the velocity cubes obtained with the SMA, overlaid with the H$_2$ and optical
images of the two classical protostellar outflows (HH 211 and DG Tau B) and the two explosive ones (Orion KL and DR 21).   For simplicity,
we have only studied the blue-shifted side of the HH 211, and red-shifted side of the DG Tau B outflow.  We argue 
that both sides of such outflows have a similar gas kinematic behaviour, as revealed in the position velocity diagrams, and 
velocity cubes of  \citet{mit1994,hir2006,palau2006,zap2015}.
However, in the case of DG Tau B it is very clear from the optical and millimeter observations that the outflow is asymmetric, with
the northwest side being more prominent, see Figure \ref{fig2}.  The latter is the side that we are studying here.  Moreover, the object exciting the
outflow in DG Tau B is associated with a T Tauri star \citep[Class I/II; ][]{mun1983}, while the source of HH 211 is related with a younger Class 0 protostar \citep{palau2006}.
In Figures \ref{fig1} and \ref{fig2}, we have included continuum compact emission from their exciting sources to clarify their locations. 
For the explosive outflows case, we have selected only three CO filaments, one in DR21 and two in Orion KL. 
These filaments are mostly isolated and so let us better trace their kinematics, see Figures \ref{fig3} and  \ref{fig4}.  
All filaments present a similar gas kinematical behaviour, {\it i.e.} well defined Hubble velocity laws, see the position-velocity CO diagrams
in \citet{zap2009,zap2013}. Hence, these filaments represent well the kinematics of the explosive outflows from a general point of view. 

In Figure \ref{fig1}, in the upper panel, we show the resulting SMA image for the blue-shifted side of the protostellar outflow HH 211 overlaid 
with the H$_2$ emission obtained from \citet{mac1994}. The blue crosses mark the position of the CO condensations at different radial velocities extracted from the
data cube. The blue crosses delineate very well the infrared H$_2$ structure and are in good agreement with the morphology revealed in
the moment zero CO maps presented in \citet{tap2012}. Our image even reveals the outflow deflection occurring on the southeast far tip of this 
blue-shifted component, and bow-shock structures across the outflow. Furthermore, the molecular outflow appears to have a double component, 
one component in the middle that is collimated (the molecular jet) and
another one more extended surrounding the molecular jet.  
This double molecular structure has already been noted in older interferometric observations \citep{gue1999} and explained in terms of a jet-driven flow.  
In this image, we have also overlaid the centimeter emission from the 
exciting object of the HH 211 outflow as mentioned before, see \citet{rod2014}. The centimeter source traces well the origin of the blue-shifted CO molecular emission. 
In the lower panel of Figure \ref{fig1}, we present the Position-Velocity Diagram (PVD) from the blue-shifted side of the flow overlaid with
the positions of the CO molecular condensations presented in the upper panel of Figure \ref{fig1}. The PVD diagram was made along the outflow axis (115$^\circ$).
In this PVD, we have also marked the position of the exciting source of HH 211 as mentioned above.   

In Figure \ref{fig2}, in the upper panel, we show the resulting SMA image for the red-shifted side of the protostellar outflow DG Tau B overlaid
with an optical HST image obtained from the HST Legacy Archive\footnote{https://hla.stsci.edu/hlaview.html}.  The crosses, as in the HH 211 outflow, 
trace the position of the CO condensations at different radial velocities extracted from the data cube, but in this case tracing red-shifted emission. 
As one can see the red crosses trace very well the entrained gas in the vicinities of the optical jet.  
The morphology of the red crosses is very similar to that obtained in the moment zero map of CO presented in \citet{zap2015}. 
We have also overlaid the millimeter continuum emission from DG Tau B obtained from \citet{zap2015}. This emission traces
the circumstellar inner disk that surrounds the exciting source of the outflow. A PVD was also extracted from the CO data cube with 
a Position Angle (P.A.) similar to that of the outflow axis (296$^\circ$). This PVD is presented in the lower panel of Figure \ref{fig2}.
In this PVD, we have also marked the position of the exciting source of DG Tau B.   

\begin{deluxetable*}{l c c c c c c c}[!]
\scriptsize
\tablecaption{Physical Parameters of the Explosive and Protostellar Outflows}
\tablehead{
\colhead{}                        &
\colhead{}                        & 
\colhead{}                        &                      
\colhead{}                        &                            
\colhead{}                        &
\colhead{Dynamical}        &
\colhead{Mechanical}            \\
\colhead{}                        & 
\colhead{Mass}                &
 \colhead{Mean Velocity}        &                      
\colhead{Momentum}       &
\colhead{Energy}              &
 \colhead{Age}                  &                           
\colhead{Luminosity}               \\
\colhead{Name}                 &
\colhead{[M$_{\odot}$]}      &
\colhead{[km s$^{-1}$]}       &
\colhead{[M$_{\odot}$ km s$^{-1}$]} &
\colhead{[10$^{45}$ erg]}   &  
\colhead{[years]}                 &                 
\colhead{[L$_{\odot}$]}       & 
}
\startdata
 {\it Explosive Outflows (One single filament) } & &  & & & & \\
Orion KL$^{a}$ &  $\geq$ 0.12 & 70  &  8.4   & 5.9  & $\sim$ 500 &  $\sim$ 93 \\
DR21$^{b}$      &  $\geq$ 0.51 & 40 & 20.5 & 8.2 &    $\sim$ 10$^4$ &  $\sim$ 6.5 \\
                & & & & & & \\
{\it Protostellar Low-mass Outflows} & & & & & & & \\
HH 211$^{e}$  (blue lobe)        &    $\geq$ 0.0015 & 20 & 0.03 & 0.006   & 3$\times$10$^3$ & 0.01 \\
DG Tau B$^{f}$        &   $\geq$ 0.0025 & 20 &  0.05 & 0.01  & 700 &  0.1 \\
               &  & &  & & & & \\  
{\it Protostellar Massive Outflows} & & & & & &\\
IRAS 18162$-$2048-NW$^{c}$   & 0.74  &  12.1 &  9 & 1.5 & 5$\times$10$^3$ &   2 \\   
IRAS 18360$-$0537$^{d}$ (blue lobe)         &  27   & 15 &  405 & 55 & 8$\times$10$^3$ & 63\\                             
\enddata
\tablecomments{\scriptsize The physical parameters of all these outflows have been obtained with an (sub)millimeter interferometer. \\
                           a). Parameters obtained from this work. The dynamical age is obtained from \citet{zap2009}. 
                           b). Parameters obtained from this work.  The dynamical age is obtained from \citet{zap2013}. 
                           c). Parameters obtained from \citet{man2013}.
                           d). Parameters obtained from \citet{qiu2012}.
                           e). Parameters obtained from this work. The dynamical age is obtained from \citet{palau2006}. 
                           f). Parameters obtained from \citet{zap2015}. 
                           }
\label{tab1}
\end{deluxetable*}

In Figure \ref{fig3}, in the upper panel, we present the image obtained by \citet{zap2009} from the explosive outflow in Orion KL,  
used here only as a reference, and a zoom into a relatively isolated zone with only a small number of filaments.  
In this zone we have selected two filaments with a similar P.A. equal to 54$^\circ$ named
Orion Filament (OF) 1 and 2.   OF1 is somehow more extended than OF2, suggesting that they may have different angles
with respect to the plane of sky.  These two filaments are composed of CO condensations at different radial velocities extracted from the
data cube. The filaments do not have clear H$_2$ counterparts, probably due to the high extinction associated
with the extended molecular ridge located in this region, see \citet{zap2011}.  In the lower panel of this Figure, we have computed the PVD of these filaments. 
This image reveals that these two filaments have in fact different inclination angles with respect to the plane of sky. 
There is also a good correspondence between the PVD and the position of the CO condensations as in the case of the HH 211 flow.
The CO gas kinematics of both filaments show the well defined Hubble laws first reported in \citet{zap2009,zap2013}. Both filaments
point to a common origin.  This origin is coincident within the errors with the position from which according to proper motion measurements the 
radio and infrared sources BN, I, and n were ejected some 500 years ago \citep{Rod05,Gom05,Gom08}. Close to the systemic velocity of the cloud in Orion
at about $+$9.0 km s$^{-1}$ and origin of the explosive flow is found the molecular expanding bubble-like outflow reported in \citet{zap2011}.   

Finally, in Figure \ref{fig4}, in the upper panel, we show the image obtained by \citet{zap2013} of the explosive outflow in DR21, 
 and a zoom into a southern region where a bright  and isolated filament is discerned.
 From this region we have selected this filament for our study. The P.A. of the filament is 215$^{\circ}$, and we have called DF1. 
 This filament DF1 like the ones we show above for Orion KL is also absent of a H$_2$ counterpart probably because of high extinction. 
 In the lower panel of this Figure, we have computed the PVD of this filament. 
 The CO gas kinematics of DF1 filament reveals the well defined Hubble law first reported in \citet{zap2013}, and also present in the filaments 
 of Orion KL. A similar expanding bubble-like structure seems evident in this PVD. This structure is similar to that mapped in Orion KL region and
 a detailed analysis of this bubble-like outflow in DR21 will be presented in a future paper.    

\section{Discussion: Explosive vs. Protostellar} 

From an observational point of view the most striking difference between the two types of outflows lies in the different numbers of 
well-defined, coherent components that make up their structures. As first revealed by SMA interferometric studies of Orion KL \citep{zap2009} 
and later by ALMA observations \citep{ball2016,ball2017}, 
an explosive outflow essentially consists of dozens to hundreds of individual filaments that extend far in space, show well defined Hubble-like velocity 
increase with distance, point towards a common center of origin and, taken together, emit the bulk of CO radiation beyond near-ambient frequencies. 
In contrast, bipolar outflows mostly present one such organized structure only, namely the collimated jet that mobilizes the otherwise chaotic, 
and incoherent outflow material.  This basic difference causes a number of important distinctions between the two classes of outflow, as discussed below.

\subsection{Spatial distributions}

The protostellar molecular outflows are bipolar in nature, and their spatial distributions of the
red- resp. blueshifted components tend to be maximally separated.    
The bipolarity of the molecular protostellar outflows comes from the fact that the ejected material from the magnetized 
disks proceed by the two poles of the protostar and the accretion flows in the equator \citep{pud2007}. 
This is clearly the case of the protostellar outflows studied here, HH 211 and DG Tau B, 
see the figures presented in \citet{gue1999,mit1994}.  
The usually large separation between the red-  and blue-shifted components in the flows is due mainly to very low probability of having a 
bipolar outflow perpendicular to the plane of the sky.  However, there is an excellent case where an outflow located perpendicular 
to the plane of the sky has been reported, namely G331.512$-$0.103 \citep{mer2013}. This bipolar
outflow shows self-absorption features at high velocities, its blueshited and redshifted sides are overlapped, 
and has very broad line wings (about 100 km s$^{-1}$).  

However, none of these geometrical effects seem to be the case for the explosive outflows where their spatial distributions of the
red- resp. blueshifted components tend to be maximally overlapping. This effect makes the explosive outflows to give the impression that they 
 are located perpendicular to the plane of the sky, and probably 
driven by wide-angle or precessing outflows from a single massive protostar. Early single dish CO images of the DR 21 
and Orion KL showed this observational effect, see Figure 10 of \citet{sch2010}, and Figure 3a of \citet{rod1999}. 
However, SMA interferometric observations revealed that this kind of flows are composed of  
dozens of collimated molecular filaments with different radial velocities and orientations that points to a single velocity 
and position center, see \citet{zap2009,zap2013}. Thus, they are mostly isotropic in all orientations (Figures \ref{fig3} and \ref{fig4}).  See the recent ALMA images 
made by \citet{ball2016,ball2017} toward the Orion KL explosive outflow.
Single dish CO observations could not resolve such collimated filaments, and it was easy to misinterpret the outflow nature. 
Even though H$_2$ images from the Orion KL and DR 21 outflows have revealed a northeast/southwest and 
northwest/southeast bipolar morphology, respectively, see Figures \ref{fig3} and  \ref{fig4}, the flows are indeed nearly isotropic in the sky
as revealed in the CO interferometric maps.    
This apparent bipolarity, noted in the H$_2$ images, could be explained as due to the high extinction of the dusty 
filaments crossing the outflows.

As mention earlier, the explosive flows are made of multiple narrow straight filament-like ejections at very different angles, while the classical protostellar 
outflows show only ejections in a narrow angle window (a central collimated molecular jet, see Figure \ref{fig1} for instance), surrounded by the 
characteristic molecular lobes.  Such narrow straight filament-like ejections are not bipolar as the  classical protostellar
outflows. The blue- and red-shifted filament-like ejections are located in a nearly isotropic distribution (see Figure \ref{fig3} and \ref{fig4}). 

\subsection{Kinematics}

There are marked and striking differences between the PVDs from the protostellar outflows and from the systems of molecular filaments within the explosive flows,
see Figures \ref{fig1}, \ref{fig2}, \ref{fig3}, and \ref{fig4}.  But the most important kinematical difference is 
the well defined Hubble flow-like increase of velocity with distance from the origin in the explosive flows versus 
the non-organized CO velocity fields in protostellar objects. We noted that the PVDs from the protostellar flows 
sometimes even reveal an anti-Hubble law, where material far from its exciting source shows low radial velocities (Figures \ref{fig1}, and \ref{fig2}), and that 
this is likely caused because the material far from the outflows is suffering deceleration by the interaction with the surrounding cloud.  
This seems not to be the case in the explosive filaments, see Figure \ref{fig3}, and \ref{fig4}.  But, maybe, some older and fainter explosive filaments
could reveal some interactions with its parental cloud using more sensitive observations.
For the case of the HH 211 flow, its PVD shown in Figure \ref{fig1}, reveals that part of the outflow appears to have a Hubble velocity law 
(it is in somehow traced by a pink line in this figure, and is material associated mostly with the molecular jet), probably generated by an internal 
working surface formed by a temporal variation in the jet velocity and observed  in many other HH objects in the optical \citep{rag1990,rag1993} 
and submillimeter regimens \citep{lee2015}.  

We note that even when this part of the flow (the collimated jet) appears to have a Hubble-law kinematics, it seems not to point to its exciting source. 
But, we stress that these two points are not entirely clear, as many other structures with different radial velocities are also overlapping with the collimated jet.  
 
Even when the outflow might seem to have an explosive morphology as in the case of Cepheus A HW2 \citep{zap2013b}, its kinematics revealed that this 
is not the case. The PVDs shown in Figure 8 of \citet{zap2013b} exhibit non-organized CO velocity fields, which are more like the 
ones displayed by the protostellar outflows. Its explosive-like morphology is probably occasioned by a precessing flow \citep{cun2009, zap2013b}.    
 
A second difference is the kinematic structure of the gas in the protostellar outflows. Both protostellar outflows (DG Tau B and HH 211) reveal ``open-fan"  structures
in their gas kinematics (see the PVD in Figures \ref{fig1}, and \ref{fig2}).  These ``open-fan" structures are composed of multiple bow-shocks at different positions 
and radial velocities, see the PVD in Figure \ref{fig1}, where we have marked the bow-shocks. The bow-shocks have a large spread of radial velocities with 
a more or less constant position (Figure \ref{fig1}), and a curved morphology.  These bow-shocks are formed by the interaction of the neutral jet ejected
from the young star with the dense molecular gas.  These structures are not observed in the PVDs of explosive outflows, see Figures \ref{fig1}, and \ref{fig2}, probably
because these flows are not driven by a stellar wind interacting for a long time with the cloud. However, we note that at least for the case of the outflow in DR21, there are
some features that could also be bow-shocks within the filaments (see Figure \ref{fig4}). This physical characteristic is in agreement with the explosive filaments being 
dense fragments of expelled material traveling into the ISM.     
 
\subsection{Energetics}

In Table \ref{tab1}, it is estimated the mass,  momentum, energy, and mechanical luminosity of the explosive filaments versus
 low-mass protostellar outflows.  We want to remark that we are comparing the energetics of the molecular filaments, not the whole explosive outflow, with
 the protostellar outflows. For comparison we have also included the energetics of some massive molecular outflows, see Table \ref{tab1}.  
We have estimated the presented physical values in Table \ref{tab1} from the equations provided in \citet{zap2015}. However, for the protostellar 
massive outflows:  IRAS 18162$-$2048-NW, and IRAS 18360$-$0537, we obtained them from \citet{man2013,qiu2012}. 
We do not find any marked difference between the values obtained for the massive protostellar outflow IRAS 18162$-$2048-NW, and the explosive filaments. 
But, if we take the protostellar outflow IRAS 18360$-$0537,
this clearly shows quite high values for the mass and momentum, compared with the explosive filaments. The outflow IRAS 18162$-$2048-NW is energized from a B-type
pre-main sequence star, while IRAS 18360$-$0537 is likely energized by O-type young star, and probably this explains the difference in the energetics. 
For the case of the low-mass flows,  there are huge differences in mass, momentum and energy compared with the explosive filaments, 
suggesting again a different nature. We note that all the physical parameters estimated here are from interferometric observations using the SMA. 
 
\subsection{Conclusions}

We have analyzed sensitive CO line SMA archival observations from the classical protostellar outflows DG TAU B and HH 211 and
the explosive ones: Orion KL and DR 21. Our conclusions are as follows.

\begin{enumerate}

\item  The CO morphology of both kinds of flows shows largely different spatial distributions 
          of the red- and blue-shifted components: maximally separated in protostellar,  
          and largely overlapping in explosive outflows. The explosive outflows consist of numerous
          narrow straight filament-like ejections with different orientations, while the classical protostellar 
          outflows show ejections in a very narrow angle window, {\it i.e.} the collimated jet.
          The filament-like ejections are not bipolar as in the case of the protostellar flows.  

\item  An in-depth study of the kinematics of the CO gas of the flows, reveals in the explosive outflows a very 
          well-defined Hubble flow-like increase of velocity with distance from the origin in contrast  to the non-organized CO velocity field in
          protostellar objects. In the protostellar outflows we commonly find well defined bow-shocks (better traced in the case of HH 211 outflow) 
          with different velocity gradients at random positions, likely produced by the interaction of the primary jet with the molecular cloud. 
          In the case of the explosive filaments, we found some bow-shocks-like structures (better observed in the DR21 outflow),
          but this needs more high angular, sensitive observations to study their nature.

\item  We report huge differences in mass, momentum and energy between the explosive filaments, and low-mass protostellar flows. 
          However, we noted that such physical parameters for the filaments in the explosive outflows  agree well with those of a B spectral-type 
          protostar molecular outflow. 
\end{enumerate}

This study thus provides new ways to distinguish between the two types of flows. With the discovery of more explosive
outflows we will unravel their nature and the physical processes that generate this phenomenon.  





\acknowledgments
LAZ, LFR, AP, and LL acknowledge the financial support from DGAPA, UNAM, and CONACyT, M\'exico.
 We  are  very  thankful  for the thoughtful suggestions of the anonymous referee that helped to improve our manuscript.\\

{\it Facilities:} \facility{The Submillimeter Array}.

\end{document}